\documentclass[twocolumn,
nofootinbib,
 amsmath,amssymb,
 aps,
 notitlepage
]{revtex4-1}

\usepackage{graphicx}
\usepackage{dcolumn}
\usepackage{bm}
\usepackage{hyperref}
\usepackage[capitalize]{cleveref}

\usepackage{todonotes}
\usepackage{amsfonts}
\usepackage{amsmath}
\usepackage{graphicx}
\usepackage{dsfont}
\usepackage{bbold}

\begin{document}

\title{Stabilizing Stuart-Landau oscillators via time-varying networks}

\author{Claudio Pereti}
\email{claudio.pereti@stud.unifi.it}
 \affiliation{Dipartimento di Fisica e Astronomia, Universit\'a di Firenze, 50019 Sesto Fiorentino, Firenze, Italy
}

\author{Duccio Fanelli}%
 \email{duccio.fanelli@unifi.it}
\affiliation{%
 Dipartimento di Fisica e Astronomia, Universit\'a di Firenze, INFN and CSDC, 50019 Sesto Fiorentino, Firenze, Italy
}

\date{\today}

\begin{abstract}
A procedure is developed and tested to enforce synchronicity in a family of Stuart-Landau oscillators, coupled through a symmetric network. The proposed method exploits network plasticity, as an inherent non autonomous drive. More specifically, we assume that the system is initially confined on a network which turns the underlying homogeneous synchronous state unstable. A properly engineered network can be always generated, which links the same set of nodes, and allows for synchronicity to be eventually restored, upon performing continuously swappings, at a sufficient rate, between the two aforementioned networks. The result is cast in rigorous terms, as follows an application of the average theorem and the critical swapping rate determined analytically.  
\end{abstract}

\maketitle


\section{Introduction}\label{sec:intro}

Synchronization is a condition of utmost coordination which is attained when a ensemble made of self-sustained oscillators evolve in unison \cite{nicolis,goldbeter,pikovsky,strogatz,kuramoto}. The simultaneous flashing of fireflies and the rhythmic applause of a large audience are often quoted as paradigmatic examples of synchronization\citep{strogatz}.  In many cases of interest, individual oscillators are localized on the nodes of a complex network\cite{barrat,osipov}. The intricate web of inter-linked connections which ultimately defines the network topology plays a role of paramount importance, when aiming at preserving the 
synchronous state \cite{barahona,arenas}. Externally imposed perturbations might in fact magnify along specific directions \cite{kori}, as reflecting the tortuous network architecture, to grow unstable in time\cite{mirollo,koseska}.  This is a possibility that should be prevented from occurring in a large plethora of applications, which rely on a degree of effective synchronization to implementing dedicated functions\cite{glass,buck}. Consider for instance the  case of distributed power grids based on renewable energy resources: suitably engineered networks  should be devised for an efficient energy deliver,  which is robust against grid disturbances such as voltage dips\cite{dorfler,dorfler1,jung}. 

Starting from these premises, the aim of this paper is to theoretically explore the possibility of achieving synchronization by acting on the network plasticity. This is accomplished by building on the methods developed in \cite{lucas}. In this latter work,  we considered a large population of non-linear, diffusively coupled oscillators and  instigated a symmetry breaking instability of a fully synchronized homogeneous equilibrium, by making the network time dependent. More specifically, we made the network to swap iteratively between two configurations, each associated to a stable regime. The network dynamics  was shown to act as a veritable non-autonomous drive by turning the examined system unstable, for a sufficiently fast pace of consecutive swappings. The condition for instability follows an application of the averaging theorem, which enables one to cast  on rigorous mathematical grounds the numerical observations\cite{lucas,julien}.  In this paper, we set to analyze the dual scenario and show that synchronization can be enforced by proper tailoring the network in time. At variance with the analysis reported in \cite{lucas}, we will here deal with a collection of coupled Stuart-Landau oscillators. These latter are characterized by a complex amplitude, the real and imaginary components evolving in time as subject to non-diagonal diffusive couplings. This setting marks another point of distinction with the analysis reported \cite{lucas}. Further, the limit cycle solution as displayed by an isolated Stuart-Landau oscillator can be written in a closed form, thus allowing for substantial analytical progress to be made.  Synchronization in time-varying random networks is also addressed in \cite{zoran},  where the links mediating the interactions are constantly rearranged. 

The paper is organized as follows. In the next section we introduce the model and characterize the stability of the associated limit cycle solution. This is achieved by expanding the perturbations superposed to the homogeneous equilibrium solution on the basis of the Laplacian operator, which governs inter-nodes exchanges. We will in particular begin by considering a simple realization of a Watts-Newman network which yields the system unstable, see Section II. We shall then generate a weighted network, different from the former, which links the same set of nodes, and engineered so as to preserve the stability of the synchronized state. Synchronicity can be eventually regained, starting from a patterned condition and by continuously swapping, with a sufficient rate, between the two aforementioned networks. In doing so, the system feels the averaged couplings among nodes, when the transition is fast enough. This intuitive observation is established on rigorous grounds in Section III. Results of the simulations which validate the theory predictions are also reported. In Section IV, the explicit calculation for the critical frequency of network modulation is provided. Finally, in Section V we sum up and draw conclusive remarks.

\section{Coupled Stuart-Landau oscillators}

Let us begin by introducing the Stuart-Landau equation:
\begin{equation}
 \frac{d W}{d t} = W - (1+ic_2){\left | W \right| ^2} W \label{2.1}
 \end{equation}
 
where $ W \in \mathbb{C}$ e $c_2 \in \mathbb{R} $. The system admits a limit cycle solution in the form $W_{LC}(t) = e^{-ic_2t}$. To characterize the stability of the 
cycle we introduce a non homogeneous perturbation in polar coordinates as:
\begin{equation}
W(t) = W_{LC}(t)[1+\rho(t)]e^{i\theta(t)}
\end{equation}

By linearizing the governing equation under the assumption of small perturbation amount yields

\begin{equation}
\frac{d}{dt}
\left(
\begin{array}{c}
\rho \\
\theta
\end{array}
\right)
=
\begin{pmatrix}
-2 & 0 \\
-2c_2 & 0
\end{pmatrix}
\left(
\begin{array}{c}
\rho\\
\theta
\end{array}
\right)
=  \mathbf{J}\left(
\begin{array}{c}
\rho \\
\theta
\end{array}
\right)
\end{equation}

where the last equality sign defines self-consistently the Jacobian matrix $\mathbf{J}$. The stability of the solution is ultimately set by the eigenvalues of $\mathbf{J}$, which can be readily computed to yield $\lambda=-2,0$. The limit cycle is therefore stable for all values of the parameter $c_2$. 

To proceed with the analysis we now take $N$  Stuart-Landau oscillators, each associated to a node of an abstract graph. The oscillators are made to interact diffusively, the nodes of the collection being linked via the edges of the graph. The complex state variable which defines individual oscillators is further decorated with a discrete index $j$,  so as to reflect the node to which the oscillator is eventually bound. The architecture of the hosting network is specified by the entries $A_{jk}$ of the weighted adjacency matrix $\mathbf{A}$. In the following we will deal with undirected networks, which in turn implies assuming $A_{jk}=A_{kj}$, for all pairs $j$ and $k$. We also assume that $A_{jk}$ can take positive or negative real values, as reflecting excitatory and inhibitory exchanges.  In formulae, we posit:

\begin{equation}
\label{coupl_SL}
\dot{W_j} = W_j - (1+ic_2)\left|W_j\right|^2W_j  + (1+ic_1)\sum_{k=1}^N A_{jk} (W_k-W_j)  
\end{equation}

with $j=1,...,N$ and where $c_1\in \mathbb{R}$ is an additional parameter of the model which sets the strength of the coupling. The above equation can be rewritten in an analogous form by introducing the discrete Laplacian operator $\mathbf{\Delta}$. To this end we define the connectivity of node $j$ as $k_j = \sum_{j}^N A_{jk}$. The elements of $\mathbf{\Delta}$ read $\Delta_{jk}=A_{jk} - k_j\delta_{jk}$. A straightforward manipulation enables us to cast equation (\ref{coupl_SL}) in the equivalent form: 

\begin{equation}
\dot{W_j} = W_j - (1+ic_2)\left|W_j\right|^2W_j  + (1+ic_1)\sum_{k=1}^N\Delta_{jk} W_k \label{2.11}
\end{equation}

which represents the discrete counterpart of the celebrated Ginzburg-Landau equation. The above system of coupled differential equations admits a trivial homogeneous solution, obtained by replicating the  limit-cycle $W_{LC}$ on each node of the network, i.e. by setting $W_j(t) = W_{LC} \qquad \forall j$, with no relative de-phasing.  The stability of the solution can be challenged by performing a perturbative calculation arrested at the linear order of approximation. In analogy with the above we set:
\begin{equation}
W_j(t) = W_{LC}(t)[1+\rho_j(t)]e^{i\theta_j(t)} \label{2.12}
\end{equation}
By inserting the ansatz  ($\ref{2.12}$) in eq. ($\ref{2.11}$) and carrying out the expansion to the leading linear order yields:

\begin{equation}
\left(
\begin{array}{c}
\rho_j' \\
\theta_j'
\end{array}
\right)
=
\begin{pmatrix}
-2 & 0 \\
-2c_2 & 0
\end{pmatrix}
\left(
\begin{array}{c}
\rho_j\\
\theta_j
\end{array}
\right)
+
\begin{pmatrix}
1 & -c_1 \\
c_1 & 1
\end{pmatrix}
\sum_{k}^N\Delta_{jk}
\left(
\begin{array}{c}
\rho_k\\
\theta_k
\end{array}
\right) \label{2.13}
\end{equation}
where the symbols $\left( ' \right)$ stands for the time derivative. To solve the above system we introduce the set of eigenvectors $\phi^{(\alpha)}$ of the Laplacian operator 
$\mathbf{\Delta}$:
\begin{equation}
\sum_{j}^N \Delta_{ij} \phi_j^{(\alpha)} = \Lambda^{(\alpha)} \phi_i^{(\alpha)} \qquad \alpha = 1,....,N \label{2.14}
\end{equation}
where $\Lambda^{(\alpha)}$ stands for the associated eigenvalues. The Laplacian operator is symmetric by definition and its eigenvalues are therefore real. It can be further proved that the eigenvalues are semi-negative, the largest eigenvalues being identically equal to zero. The eigenvectors $\phi^{(\alpha)}$ forms an orthonormal basis, which can be exploited for reducing the complexity of the linear system $(\ref{2.13})$. To this end we expand the perturbation $\rho_j$ e $\theta_j$ on the basis $\{\phi^{\alpha}\}$
\begin{equation}
\left(
\begin{array}{c}
\rho_j \\
\theta_j
\end{array}
\right)
= \sum_{\alpha=1}^N
\left(
\begin{array}{c}
\rho^{(\alpha)} \\
\theta^{(\alpha)}
\end{array}
\right)
e^{\lambda t}
\phi_j^{(\alpha)}  \label{2.15}
\end{equation}

By inserting the above relation in eq. ($\ref{2.13}$) and making use of eq. ($\ref{2.14}$), one gets:
\begin{equation}
\left(
\begin{array}{c}
\rho^{(\alpha)} \\
\theta^{(\alpha)}
\end{array}
\right)\lambda =
\begin{pmatrix}
-2 & 0 \\
-2c_2 & 0
\end{pmatrix}
\left(
\begin{array}{c}
\rho^{(\alpha)}\\
\theta^{(\alpha)}
\end{array}
\right)
+
\begin{pmatrix}
1 & -c_1 \\
c_1 & 1
\end{pmatrix}
\Lambda^{(\alpha)}
\left(
\begin{array}{c}
\rho^{(\alpha)}\\
\theta^{(\alpha)}
\end{array}
\right) \label{2.16}
\end{equation}
To allow for non trivial solution of system ($\ref{2.16}$) we ought to require:
\begin{equation}
det
\begin{pmatrix}
-2+\Lambda^{(\alpha)} - \lambda & -c_1\Lambda^{(\alpha)}\\
-2c_2+c_1\Lambda^{(\alpha)} & \Lambda^{(\alpha)} - \lambda
\end{pmatrix}
=0
\label{matrix}
\end{equation}
The fate of the imposed perturbation can be assessed by estimating $\lambda$, the eigenvalues of the Jacobian matrix modified with the inclusion of spatial couplings, which reflect back in the terms proportional to $\Lambda^{(\alpha)}$. Solving the characteristic polynomial from (\ref{matrix}) and selecting the eigenvalue with the largest real part, we obtain:
\begin{equation}
\lambda_{max}(\Lambda^{(\alpha)}) = \Lambda^{(\alpha)} -1 + \sqrt{1+2c_1c_2\Lambda^{(\alpha)} - c_1^2\Lambda^{(\alpha)2}} \label{2.19}
\end{equation}
which is often referred to as the dispersion relation.  If $\lambda_{max}$ is positive over a subset of $\Lambda^{(\alpha)}$, the perturbation grows exponentially in time, breaking the symmetry of the initial homogeneous state. At variance,  when  $\lambda_{max} \le 0$, for all $\Lambda^{(\alpha)}$, the perturbation fades away and the system eventually regains the fully synchronized state. The conditions for the onset of the instability can be indeed condensed in a compact formula. Imagine  $\Lambda^{(\alpha)}$ to behave as continuum variable and expands relation (\ref{2.19}) for small values of $\Lambda^{(\alpha)}$. One readily gets $\lambda_{max} \simeq (1+c_1c_2) \Lambda^{(\alpha)}$\cite{cencetti}. Since $\Lambda^{(\alpha)}<0$, by definition, it is sufficient to require $(1+c_1c_2) <0$ for the instability to develop. This latter condition defines the generalization of the Benjamin-Feir instability to a Ginzburg-Landau equation defined on a network \cite{nakao,dipatti}. On the other hand we should make sure that there is at least one eigenvalue of the discrete Laplacian operator which falls in the region where $\lambda_{max}$ is positive (it is immediate to conclude that $\lambda_{max} \rightarrow -\infty$, when  $|\Lambda^{(\alpha)}| \rightarrow \infty$). This is achieved by requiring 
$\delta \Lambda < \frac{ -2(c_1c_2+1)}{1+c_1^2}$, where $\delta \Lambda$ stands for the spectral gap, i.e.  the difference in magnitude between the first and second eigenvalue.

\begin{figure}[htbp]
\includegraphics[scale=0.5]{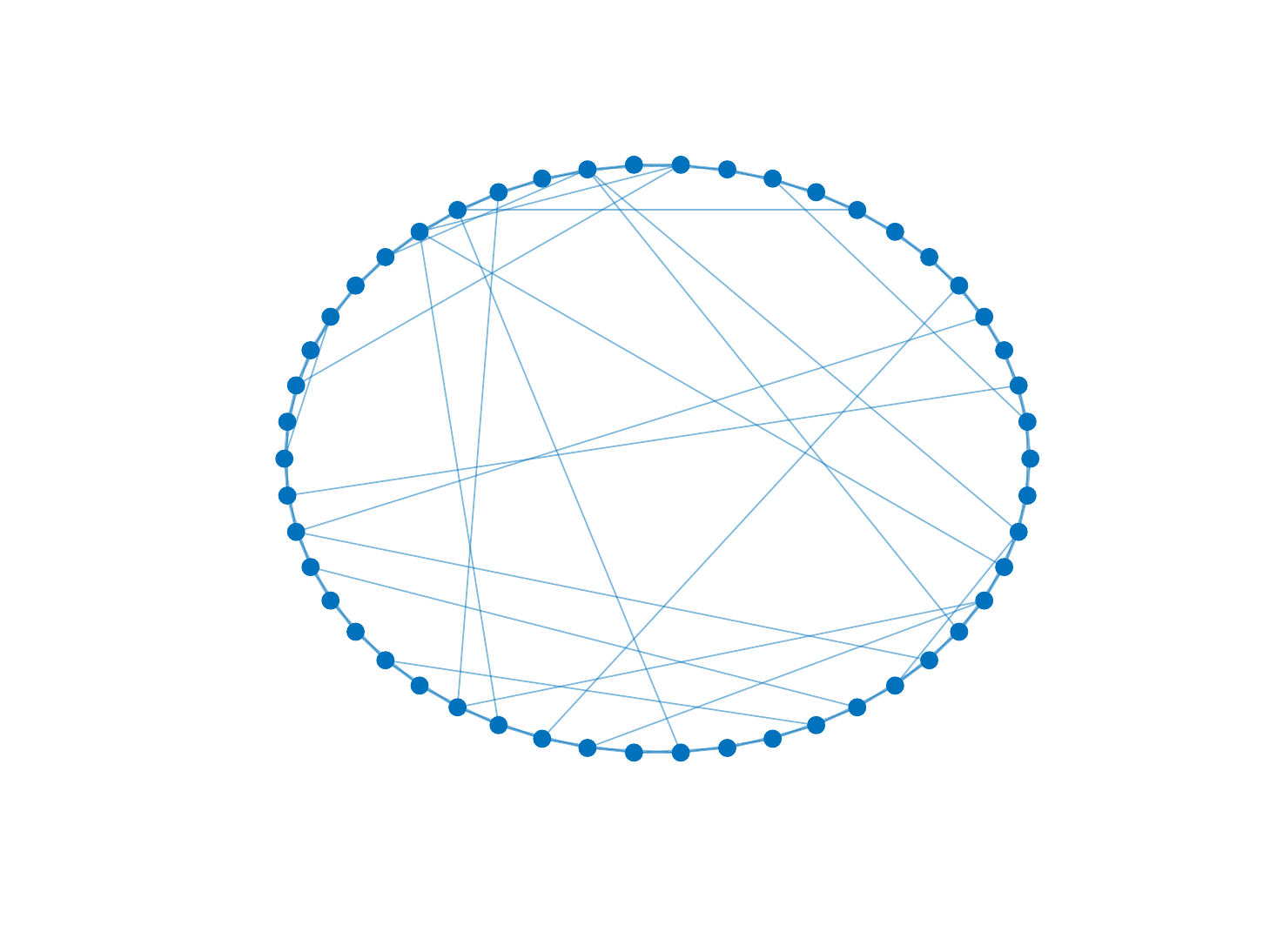}
\caption{An undirected network is generated with $N=50$ by means of the Watts-Newman prescription.}
\label{Fig1}
\end{figure}

\begin{figure}[htbp]
\includegraphics[scale=0.5]{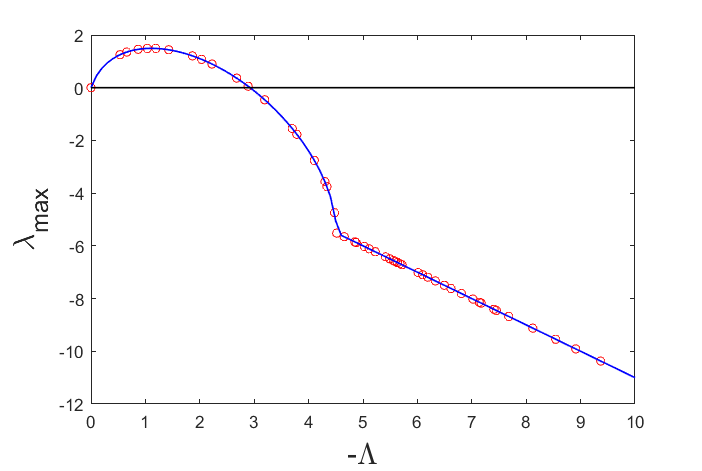}
\caption{The dispersion relation is displayed for  $c_1 = -1.8$ and $c_2 = 4$, and by assuming the network depicted in Fig.  \ref{Fig1} to provide the backbone for the model. The symbols stand for the discrete dispersion relation as obtained for the system of Stuart-Landau oscillators coupled through the network. The solid line refers to the continuous dispersion relation which is eventually attained when letting   $\Lambda^{(\alpha)}$  to change continuously within its domain of definition.}
\label{Fig2}
\end{figure}

\begin{figure}[htbp]
\includegraphics[scale=0.5]{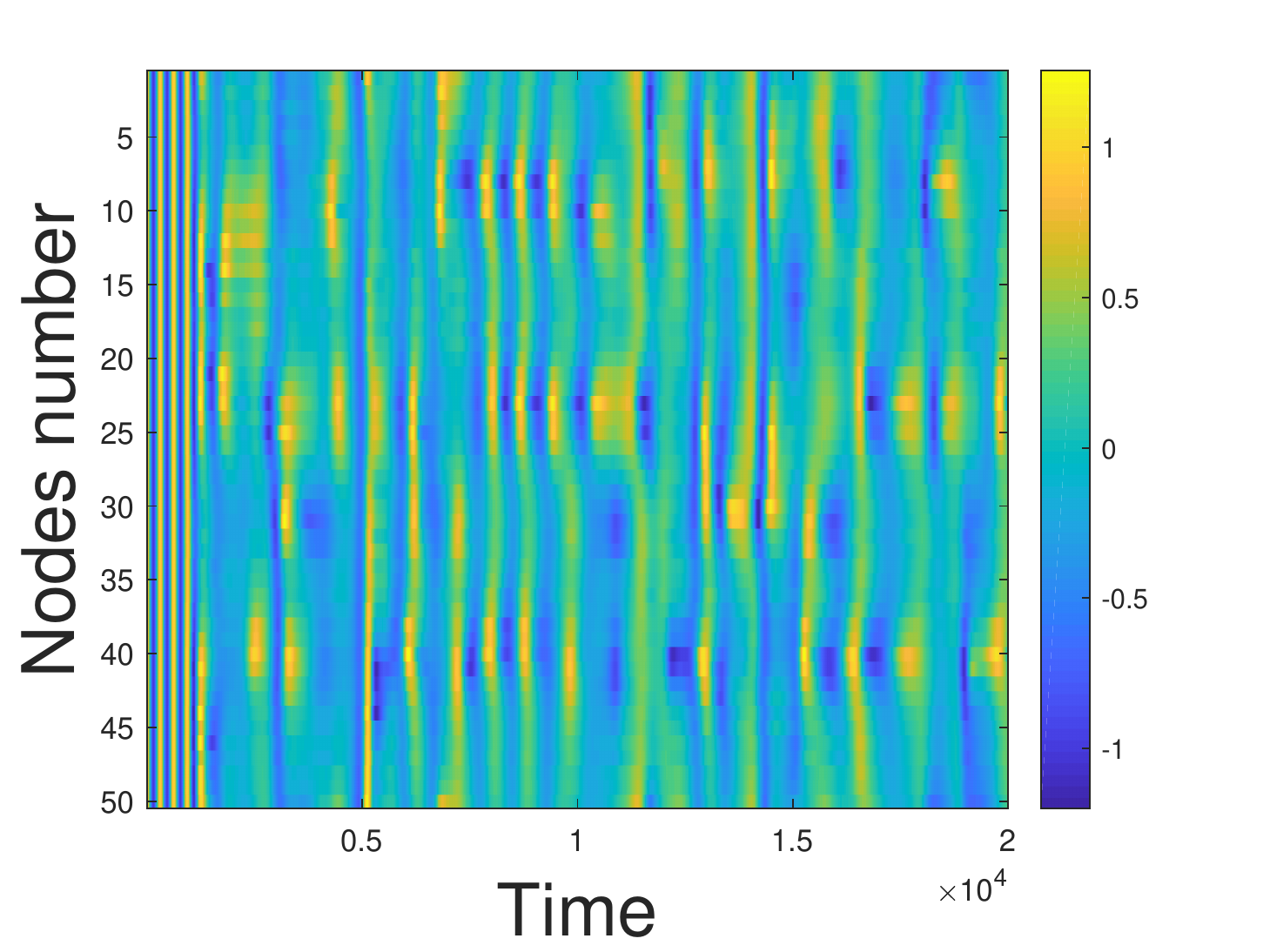}
\caption{Time evolution of the system of coupled Stuart-Landau oscillators for the setting discussed in Fig. \ref{Fig2} and upon injection of a tiny perturbation at $t=0$. The synchronicity gets broken and the system evolves towards a patterned distribution for the norm of the complex field $W_j$.}
\label{Fig3}
\end{figure}

To illustrate the factual implications of the conclusions reached above, we generate a symmetric network with the Watts-Newman algorithm\cite{newman} and label with $\bf{A}_1 $ its associated adjacency matrix. The network is depicted in Fig. \ref{Fig1}. We then calculate the dispersion relation for the collection of Stuart-Landau oscillators distributed on the node of the network and therein interacting as dictated by eq. (\ref{2.11}). The obtained dispersion relation is displayed in Fig. \ref{Fig2}, with circles. The solid line is a guide for the eye and it is formally recovered when $\Lambda^{(\alpha)}$ is made to change continuously within its domain of definition. The symbols protrude in the region of positive $\lambda_{max}$, thus implying that the homogeneous synchronous solution is unstable to external perturbation. In Fig. \ref{Fig3} the dynamical evolution of the system is represented, by plotting the norm of $W_j$ against time, for all nodes and with an appropriate color code.  The oscillators are initially in phase. The injection of a tiny perturbation breaks however the symmetry of the initial solution driving the emergence of patterned motifs. Starting from this setting, we here aim at designing a suitable control strategy to enforce stability by making the adjacency matrix time dependent.

\section{Evolving the Stuart-Landau oscillators on a time dependent network.}

As stated earlier, our aim is to make the embedding network time dependent so as to eventually achieve a dynamical stabilization of the synchronous regime. With this objective in mind, we set here to adapt to the present context the analysis discussed in \cite{lucas}. It is important to emphasise that, in the current framework, the coupling between oscillators is non diagonal, at variance with the schemes investigated in \cite{lucas}. As the network evolves in time, we rewrite eq. (\ref{2.11}) in the form:

\begin{equation}
\dot{W_j} = W_j - (1+ic_2)\left|W_j\right|^2W_j  + (1+ic_1)\sum_{k=1}^N\Delta_{jk}(t/\epsilon) W_k  \label{31}
\end{equation} 

with $j=1,...,N$ and where now the Laplacian operator depends on time, as the adjacency matrix does. The parameter $\epsilon$ sets the time-scale of the Laplacian dynamics. We will be in particular focusing on the case of interest in which the network is periodically rearranged, by iterating in time periodic swaps between two static networks: the one introduced in the preceding section (and which drives the dynamics of the system unstable) and a suitably designed network, engineered so as to enforce stabilization. We will label with $T_s$ the period of the network modulation as obtained for $\epsilon=1$. When multiple swaps between two static networks are produced over one period $T_s$, $\epsilon$ sets the frequency of the blinking. 
In the final part of the paper we will prove, that stabilization of the synchronous state can be produced by introducing a suitable partner network to $\bf{A}_1 $ and making $\epsilon$ sufficiently small. 
The remaining part of this section, is devoted to framing within the current setting the generalization of the averaging theorem discussed in \cite{lucas}.  The critical value of 
$\epsilon$ above which stability is regained, can be  computed as a byproduct of the analysis.

To make the equations compact  introduce the $2N$ elements vector $\vec{x}= (\rho_1,...,\rho_N,\theta_1,...\theta_N)^T$ . Define then the $N \times N$ matrix 
$\bf{F}=\begin{pmatrix}
-2\mathbb{1} & \bf{0} \\
-2c_2\mathbb{1} & \bf{0}
\end{pmatrix}$ 
and label $\bf{\Delta}=\begin{pmatrix}
\Delta & -c_1\Delta \\
c_1\Delta & \Delta
\end{pmatrix}
$.
Hence, the linear equations (\ref{2.13}) for the system defined on a time dependent network, as in light of the above discussion, 
takes the form:
\begin{equation}
\dot{\vec{x}} = (\mathbf{F} + \mathbf{\Delta}(t/\epsilon))\vec{x} \label{3.5}
\end{equation}
We further define the {\it averaged} system  as: 
\begin{equation}
\dot{\vec{y}} = (\mathbf{F} +\mathbf{\left\langle \Delta\right\rangle})\vec{y} \label{3.6}
\end{equation}
where the average Laplacian reads $\left\langle \mathbf{\Delta} \right\rangle = 1/T_s \int_{0}^{T_s}\Delta dt$.
Next step in the analysis is to rescale time as  $\tau = t/\epsilon$.  Eq. (\ref{3.5}) transforms into:
\begin{equation}
\vec{x}' = \epsilon(\mathbf{F} + \mathbf{\Delta}(\tau))\vec{x}
\end{equation}
where $\vec{x}'$ denotes the derivative with respect to the rescaled time variable $\tau$. In complete analogy, the averaged system writes:
\begin{equation}
\vec{y}' = \epsilon(\mathbf{F} + \left\langle\bf{\Delta}\right\rangle) \vec{y}
\end{equation} 

Label $\vec{y}$ the solution of the (partially) averaged system and with $\vec{x}$ the solution obtained for the system in its original, fully time dependent, formulation. In the following we will show that $\vec{y} - \vec{x} = \mathcal{O}(\epsilon)$, for $\epsilon \rightarrow 0$, and over times which scale as $1/\epsilon$,  provided $\vec{y}(0) = \vec{x}(0)$. To reach this conclusion, and following the path outlined in \cite{lucas}, we will consider a more general framework of the type: 
\begin{equation}
\dot{\vec{x}} = \epsilon f_1(\vec{x},\epsilon t) + \epsilon f_2(\vec{x},t) \label{3.9}
\end{equation}
where  $\vec{x}\in \mathbb{R}^{\Omega}$, $f_1(\vec{x},t) \in \mathbb{R}^{\Omega} \times \mathbb{R}\rightarrow \mathbb{R}^{\Omega}$ is $T$-periodic in $t$ and $f_2(\vec{x},t) \in \mathbb{R}^{\Omega} \times \mathbb{R}\rightarrow \mathbb{R}^{\Omega}$ is $T_s$-periodic in $t$. Assume that $f_2$ e $f_1$ and their derivatives are 
Lipschitz functions of the variable $\vec{x}$. Notice that $f_1(\vec{x},\epsilon t)$ is $T/\epsilon$-periodic.

Define: 
\begin{equation}
 \vec{u}(\vec{x},t) = \int_0^t ds[f_2(\vec{x},s) - \left\langle f_2 \right\rangle] \label{3.10}
\end{equation}
where $\left \langle f_2\right\rangle = 1/T_s \int_0^{T_s} f_2(\vec{x},t)dt$ is the average of $f_2$ over its own period. Introduce the near-identity  transformation :
\begin{equation}
\vec{x}(t) = \vec{z}(t) + \epsilon \vec{u}(\vec{z}(t),t) 
\end{equation}
which yields:
\begin{equation}
\dot{\vec{x}} = \dot{\vec{z}} + \epsilon \frac{\partial{\vec{u}}}{\partial{\vec{z}}} \dot{\vec{z}} + \epsilon \frac{\partial{\vec{u}}}{\partial{t}}
\end{equation}
By accounting for the definition of  $\vec{u}$ (see eq. $\ref{3.10}$) one gets: 
\begin{equation}
\frac{\partial{\vec{u}}}{\partial{t}}(\vec{z},t) = f_2(\vec{z},t) - \left\langle f_2\right \rangle
\end{equation}
and by inserting in (\ref{3.9}) yields:
\begin{equation}
\mathbf{\Gamma} \dot{\vec{z}} = [\mathbb{1} + \epsilon \frac{\partial{\vec{u}}}{\partial{ \vec{z}}}]\dot{\vec{z}} = \epsilon[f_1(\vec{z} + \epsilon \vec{u},\epsilon t) + f_2(\vec{z} + \epsilon \vec{u},t) - f_2(\vec{z},t) + \left\langle f_2 \right\rangle]
\end{equation}
By recalling that  $f_2$ is a Lipschitz function and �that $\vec{u}$ is bound, we obtain
\begin{equation}
 \left || f_2(\vec{z} + \epsilon \vec{u},t) - f_2(\vec{z},t) \right || < L \epsilon \left || \vec{u}(\vec{z},t) \right || < L \epsilon M
\end{equation}
where $L$ and $M$ are positive constants. Hence:
\begin{equation}
\mathbf{\Gamma} \dot{\vec{z}} = \epsilon f_1(\vec{z} + \epsilon \vec{u},\epsilon t) + \epsilon \left\langle f_2 \right\rangle + \mathcal{O}(\epsilon^2) \simeq  \epsilon f_1(\vec{z},\epsilon t) + \epsilon \left\langle f_2 \right\rangle
\end{equation}

In general $\mathbf{\Gamma}$ is not invertible, but the identity is and, by continuity, one can also invert any matrix sufficiently close to it. Hence, there exists a 
critical value of  $\epsilon^*<<1$ such that $\mathbf{\Gamma}$ is invertible in the range $0<\epsilon<\epsilon^*$. Up to $\mathcal{O}(\epsilon)$ corrections, we can therefore write:

\begin{equation}
\mathbf{\Gamma}^{-1} = [\mathbb{1}- \epsilon \frac{\partial{\vec{u}}}{\partial{\vec{z}}}]
\end{equation} 
or, equivalently:
\begin{equation}
\dot{\vec{z}} \simeq \epsilon [f_1 + \left\langle f_2 \right\rangle]
\end{equation}

We are therefore in the position to conclude that system (\ref{3.9}) behaves as its \emph{partially averaged} counterpart, for times of the order $1/\epsilon$, $\epsilon$ being a small parameter.  The same conclusion holds true for systems (\ref{3.5}) and (\ref{3.6}) over times $\mathcal{O}(1)$,  as it readily follows by recalling the definition of $\tau$.

The above general result can be invoked to hold in the simplified framework that proves here of interest.  Owing to the particular structure of the discrete Ginzburg-Landau equation, in fact, the only term which senses a non-autonomous drive is that ruling the inter-nodes couplings. We recall that for the case at hand, this takes the form of a $2N \times 2N$ matrix proportional to the Laplacian operator $\mathbf{\Delta}$, periodic with period $T_s$. When considering, however, a set of general non linear oscillators mutually inter-tangled via a Laplacian coupling, one should deal with a time dependent reaction part, this latter being characterized by a different periodicity of that associated to the coupling term. For this reason, and aiming at the broader picture, we have here decided to carry out the analysis for the general setting where $f_1$ is also time dependent.  
  
Back to our problem, from eq. (\ref{3.9}) we can eventually recover eq. (\ref{3.5}), by performing the following manipulations:  $t \rightarrow t/\epsilon $,  $\vec{x} \rightarrow \vec{x}$, $f_1 \rightarrow \mathbf{F}\vec{x}$, $f_2(\vec{x},t) \rightarrow \mathbf{\Delta}\vec{x}$ e $\Omega = 2N$. The averaged system admits a synchronous states whose stability can be assessed by performing a linear stability analysis which is completely analogous to that detailed in the preceding section. In this case, the perturbation needs to be expanded on the basis formed by the eigenvectors of the average Laplacian $\langle \bf{\Delta} \rangle$ to eventually yield a dispersion relation which is formally identical to eq. (\ref{2.19}), except for the fact that now $\Lambda^{(\alpha)}$ stand the eigenvalues of $\langle \bf{\Delta} \rangle$. In the following, we will show that a second static network, characterized by the adjacency matrix $\mathbf{A_2}$ can be always generated, that yields a stable dynamics for the {\it average} system, when combined to network characterized by the unstable topology $\mathbf{A_1}$. If the swapping between the two networks (the initial network supposed to yield an unstable dynamics and the one designed to enforce stability) is fast enough, namely if $\epsilon$ is sufficiently small, the system behaves as its average counterpart, at short times.  This allows for the examined system, subject to time-dependent couplings, to stably regain the synchronized regime, a prediction of the theory that is corroborated by numerical simulations, as we shall herefter see \footnote{Indeed, the averaging theorem ensures the stability of the time-modulated system for times $\mathcal{O}(1)$. The fact that robust synchronicity is permanently secured is numerically confirmed by integrating the system for extremely long times.}. In the following, we will describe the strategy to generate the second network to be paired to the first, so as to drive the system stable against the injection of external perturbations.

We begin with the network characterized by the matrix $\bf{A}_1$, as introduced above which yields the Laplacian $\bf{\Delta}_1$. By hypothesis the system is unstable, when defined on the latter network. To enforce stabilization we will introduce a second network, characterized by the adjacency matrix $\bf{A}_2$, and make the system to alternate, sufficiently fast, between the two static networks. This process results in the -- time dependent -- adjacency matrix $\mathbf{A}$ which is defined as follows over one period $T_s$: 

\begin{equation}
\mathbf{A}(t) =
\begin{cases}
\mathbf{A}_1 \qquad if \qquad t\in [0,T_s \gamma[ \\
\mathbf{A}_2 \qquad if \qquad t\in [T_s \gamma,T_s[
\end{cases}
\label{defA}
\end{equation}

where $\gamma$ (resp. $1- \gamma $) represents the fraction of  $T_s$ that the system spends on the network characterized by matrix $\mathbf{A}_1$ (resp. $\mathbf{A}_2$). Recall that the period, and hence the time duration of each phase, can be further tuned via the control parameter $\epsilon$.  The adjacency matrix $\bf{A}_2$ yields the Laplacian  $\bf{\Delta}_2$. The average Laplacian reads consequently  $\langle \bf{\Delta}\rangle = \gamma \bf{\Delta}_1 + (1-\gamma) \bf{\Delta}_2$. 

Imagine, as a starting point, that we aim at modifying the $k$-th eigenvalue of  $\bf{\Delta}_1$, which reads $\Lambda^{(k)}$. Our goal is to set up an average Laplacian $\langle\bf{\Delta}\rangle$, where $\Lambda^{(k)}$ has been replaced into, say,  $\mu$ while the other eigenvalues are left unchanged. This is achieved by defining $\bf{\Delta}_2$ as

\begin{eqnarray}
{\bf{\Delta}}_2 &=& \Lambda^{(1)} \psi^{(1)} \psi^{(1) T} + \Lambda^{(2)} \psi^{(2)} \psi^{(2) T} + ...\\ \nonumber
 &+&\eta \psi^{(k)} \psi^{(k) T} + .. +\Lambda^{(N)} \psi^{(N)} \psi^{(N) T}
\end{eqnarray}

and choosing  $\eta$ in such a way that the condition $\mu =  {\gamma} \Lambda^{(k)} + (1-\gamma) \eta$ is eventually met. In compact notation:
 \begin{equation}
  {\bf{\Delta}}_2 = {\bf{\Delta}}_1 + (\eta - \Lambda^{(k)}) \psi^{(k)} \psi^{(k) T}
\end{equation}
The average Laplacian, in its spectral decomposition, takes therefore the form:
\begin{eqnarray}
  \langle {\bf{\Delta}} \rangle &=& \Lambda^{(1)} \psi^{(1)} \psi^{(1) T} + \Lambda^{(2)} \psi^{(2)} \psi^{(2) T} + ...\\ \nonumber 
  &+&((1-\gamma)\eta + \gamma\Lambda^{(k)}) \psi^{(k)} \psi^{(k) T} + .. +\Lambda^{(N)} \psi^{(N)} \psi^{(N) T}
\end{eqnarray}

and it is straightforward to verify that the eigenvalue relative to the eigenvector $\psi^{(k)}$ takes the desired value:
\begin{equation}
   \langle {\bf{\Delta}} \rangle \psi^{(k)} =  \left({\gamma} \Lambda^{(k)} +(1-{\gamma}) \eta \right) \psi^{(k)} = \mu \psi^{(k)}
\end{equation}
since $\psi^{(\alpha )T} \psi ^{(k)} = 0$ for $\alpha \neq k$ and $\psi^{(k) T} \psi ^{(k)} = 1$.\\

Once the average Laplacian has been produced, for a specific choice of $\gamma$, one can go back to calculating the sought matrix $\bf{A}_2$, $\bf{A}_1$ being a priori assigned.

\begin{figure}[htpb]
\includegraphics[scale=0.5]{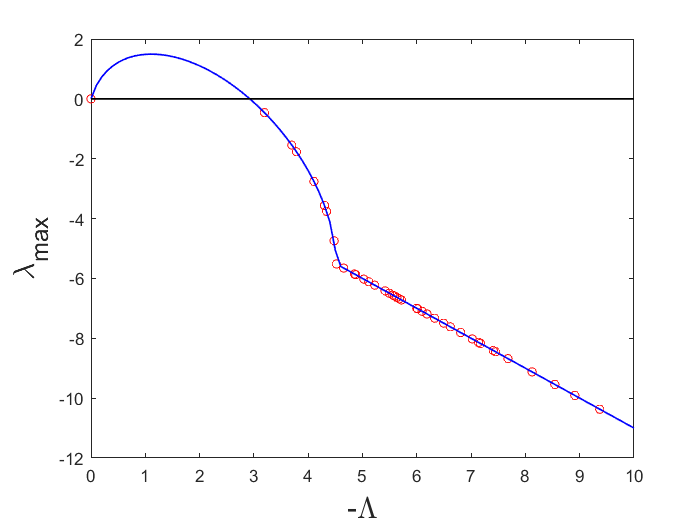}
\centering
\caption{Dispersion relation for the Stuart-Landau oscillators coupled via the symmetric Laplacian $\bf{\Delta}_2$, generated via the procedure illustrated in the main text. Symbols refer to the discrete dispersion relation, while the solid line stands for the homologous curve obtained for the system defined on a continuous support. Here, $c_1 = -1.8$ and $c_2 = 4$.}
\label{Fig_reldisp2}
\end{figure}

\begin{figure}[htpb]
\includegraphics[scale=0.5]{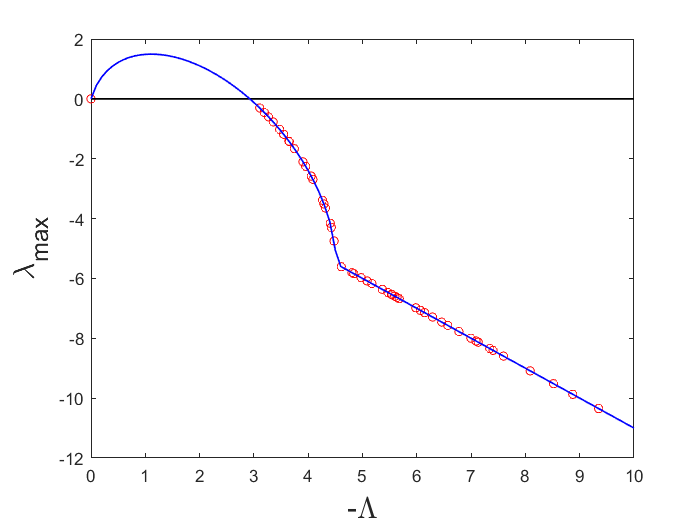}
\centering
\caption{Dispersion relation for the Stuart-Landau oscillators coupled via the average Laplacian $\left\langle \bf{\Delta} \right\rangle$ , with ${\gamma}=0.5$. Symbols refer to the discrete dispersion relation, while the solid line stands for the homologous curve obtained for the system defined on a continuous support. Here, $c_1 = -1.8$ and $c_2 = 4$ }
\label{Fig_reldisp3}
\end{figure}

The above procedure easily generalizes to the case of interest where an ensemble made of eigenvalues are to be changed. In particular, for the instability to be controlled, we need to force the eigenvalues of the average Laplacian, to be larger, in absolute value, of the critical  threshold $s = -2(1+c_1c_2)/(1 + c_1^2)$. Let us begin by ordering the eigenvalues of the Laplacian $\bf{\Delta}_1$ from the largest (by definition the one identically equal to zero) to the smallest. Assume, without losing generality, that the first $m$ eigenvalues which follows in the ranking $\Lambda^{(1)}=0$, falls in the region of instability.  We then apply the above reasoning to act on these latter $m$ eigenvalues of $\bf{\Delta}_1$, which are ultimately responsible for driving the system unstable. This is achieved by setting  $\mu(k) = {\gamma} \Lambda^{(k)} + (1-{\gamma}) \eta(k)$ with $k = 2,...,m+1$ and selecting $\eta(k)$ in such a way that $|\mu(k)| > s$, for  $k = 2,...,m+1$. 
In Fig. \ref{Fig_reldisp2}, the dispersion relation for the Stuart-Landau oscillators coupled via the newly generated Laplacian  $\bf{\Delta}_2$ is plotted. The system coupled via network 
$\bf{A}_2$ is therefore stable. More importantly the system is also stable (by construction) when defined on the averaged  Laplacian $\left\langle\bf{\Delta}\right\rangle$, as it can be appreciated by direct inspection of Fig. \ref{Fig_reldisp3}.

To prove the validity of the theory we construct the following numerical experiment. We initially confine the system on the network specified by the adjacency matrix $\bf{A}_1$. The system is started from a synchronous solution and a tiny non homogeneous perturbation added. As predicted by the theory (see dispersion relation depicted in Fig. \ref{Fig2}), the synchrony breaks apart and a patterned configuration for the norm of the complex variable $W_j$ develops. At this point, we make the system to swap continuously between the network specified by the adjacency matrix $\bf{A}_1$ and that associated to matrix $\bf{A}_2$ (which can be straightforwardly computed from  Laplacian $\bf{\Delta}_2$), with a proper choice of the parameter $\epsilon$ \footnote{The obtained network is fully connected, with positive and negative weights. It can be a posteriori trimmed by eradicating unimportant links, namely those links which do not substantially alter the dispersion relation calculated for the average network.}. The perturbation, which has materialized in the patchy distribution from the amplitude of the complex variable across nodes, gets eventually re-absorbed and the system converges back to the synchronized solution. Indeed, for a sufficiently small choice of $\epsilon$, the system behaves as if it was sensing the average network associated to the discrete Laplacian $\left\langle  \bf{\Delta} \right\rangle$ which yields the stable dispersion relation plotted in Fig. \ref{Fig_reldisp3}. In the following Section, we will elaborate on a viable strategy to determine the critical threshold in $\epsilon$. The analysis follows closely the approach employed in the Supplementary Material annexed to reference \cite{julien}.

\begin{figure}[tbp]
\includegraphics[scale=0.6]{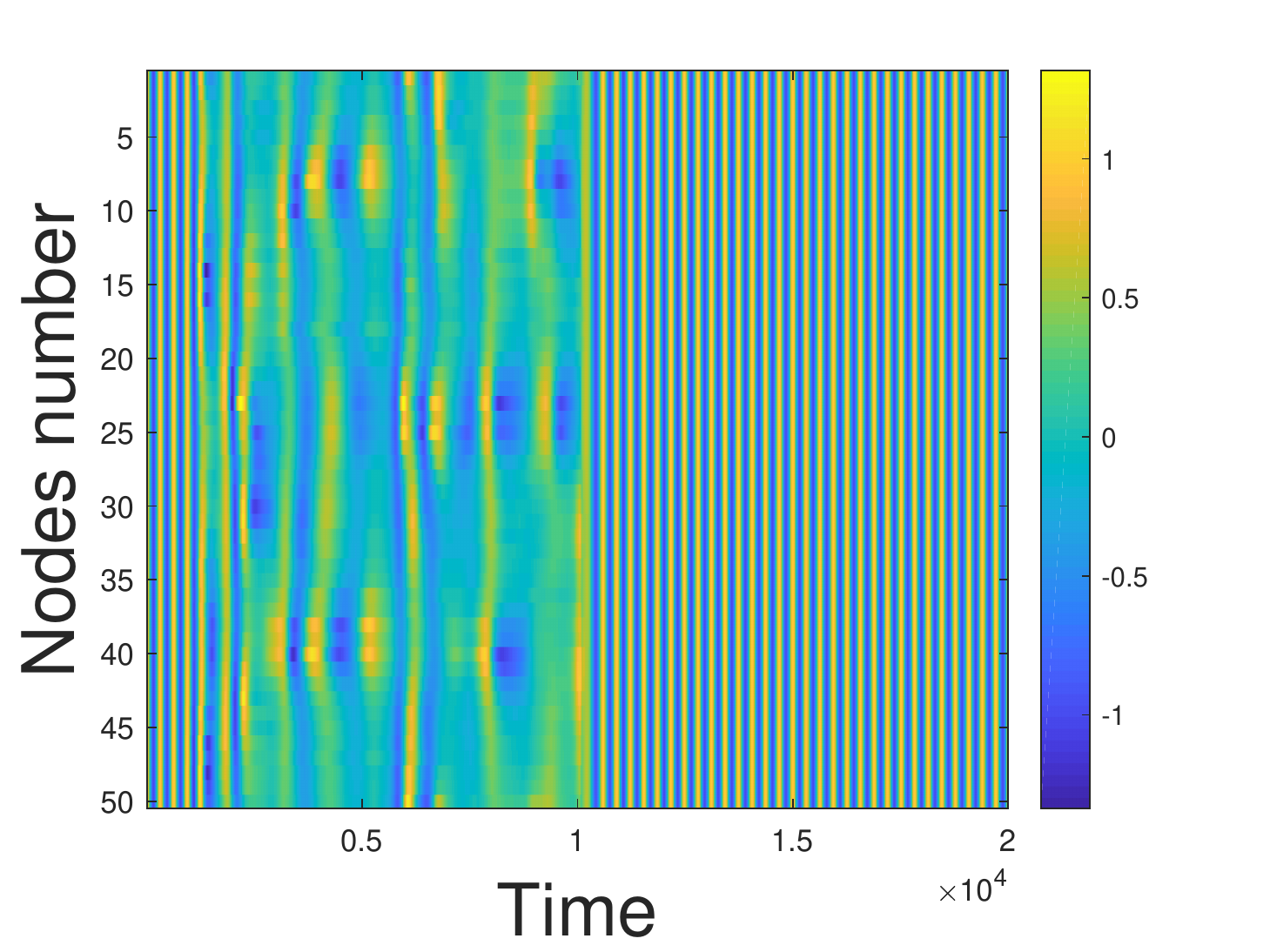}
\centering
\caption{Time evolution of the system of coupled Stuart-Landau oscillators: the modulus of $W_j$ is plotted against time. Initially the system is made to interact via the network specified by the adjacency matrix $\bf{A}_1$ and a small non homogeneous perturbation is inserted to break the synchronicity of the solution. When the patterned motifs are developed, we start swapping in between networks $\bf{A}_1$ and $\bf{A}_2$, as assumed in (\ref{defA}). Here, ${\gamma}=0.5$. The dynamics of the system follows the ruling equations (\ref{31}), with $\epsilon=0.01$. The perturbation, although already developed in its non linear stage, gets eventually re-absorbed and the system converges back to its fully synchronized condition. }
\label{simmd}
\end{figure}

 \section{Determining the critical value of $\epsilon$.}

As time progresses, the network swaps continuously between two configurations respectively characterized by the adjacency matrices $\bf{A}_1$ and $\bf{A}_2$. In the following, we shall adopt the alternative, although equivalent, notation ${\bf{A}}^{[k]}$ con $k=1,2$ and proceeds similarly for ${\bf{\Delta}}$. The changes in the networks reflects back in a modulation of the Laplacian matrix: 
\begin{equation}
{\bf{\Delta}}(t) = {\bf{\Delta}} ^{[\sigma(t)]}
\end{equation}
where $\sigma(t)$ is a piecewise constant function such that  $\sigma:[0,\infty)\rightarrow K$ with $K = {1,2}$. The function $\sigma$ enables one to associate to each time its corresponding network.  More explicitly, we have  $\sigma (t) = 1$ if $mod(t/T_s) \in [0,{\gamma})$ e $\sigma(t) = 2$ otherwise. By making use of the introduced notation, the governing system takes the form:

\begin{equation}
  \dot{\vec{x}} = \mathbf{M}_{\epsilon}^{[\sigma(t)]}\vec{x}
\end{equation}

where ${\bf{M}}_{\epsilon}(t) = \mathbf{F} + {\mathbf{\Delta}}^{[\sigma(t/\epsilon)]}$.

Let $0=t_0<t_1<t_2.....$ to identify the times for which $\sigma(t)$ has a discontinuity and define $\tau_k$ such that $t_k + \tau_{k+1} = t_{k+1}$,  $\forall k \in \mathbb{N}$. The function $\sigma(t)$ is constant in  $[t_k,t_{k+1})$, and hence the above system writes as:  

\begin{equation}
 \vec{x}(\epsilon t_{k+1}) = \exp(\epsilon \tau_{k+1} \mathbf{M}_\epsilon^{[\sigma(\epsilon t_{k})]}) \vec{x}(\epsilon t_{k})
\end{equation}

with $k \in \mathbb{N}$. We have therefore obtained a time discrete representation of the dynamics. This latter is stable provided

\begin{equation}
\rho \lbrace \exp(\epsilon \tau_k \mathbf{M}_\epsilon^{[\sigma(\epsilon t_k)]}, k \in \mathbb{N}\rbrace < 1
\end{equation}

where $\rho(\cdot)$ stands for the radial spectrum. The above expression simplifies if we assume that the function $\sigma$ admits a finite set of $n$ discontinuities and that it is $T_s$-periodic, with $T_s = \sum_{i=1}^n \tau_i$. The solution at time $\epsilon m T_s$, reads
\begin{equation}
\vec{x}(\epsilon m T) = (\mathbf{Q}_\epsilon(\epsilon T))^m \vec{x}(0)
\end{equation}
where the operator which sets the evolution over one period writes:
\begin{equation}
\mathbf{Q}_\epsilon (\epsilon T) = \prod _{k=1} ^n \exp(\epsilon \tau_k \mathbf{M}_\epsilon ^{[\sigma(\epsilon t_{k-1})]}) = \prod _{k=1} ^n \exp(\epsilon \tau_k \mathbf{M} ^{[\sigma(t_{k-1})]})
\end{equation}
and where usee has been made of ${\bf{M}}_\epsilon ^{[\sigma(t)]}={\bf{M}}^{[\sigma(t/\epsilon)]} $, and thus  ${\bf{M}}_\epsilon ^{[\sigma(\epsilon t)]}={\bf{M}}^{[\sigma(t)]} $. 
The stability is hence achieved provided $\epsilon<\epsilon^{\ast}$, with 

\begin{equation}
\epsilon^{\ast} = \min \lbrace \epsilon > 0 : \forall p \geq \epsilon, \rho(\mathbf{Q}_p(pT)) >  1 \rbrace
\end{equation}

A straightforward calculation returns $\epsilon^* = 2.6330$, for $c_2= 4$ e $c_1 = -1.8$, and estimate which agrees perfectly against numerical simulations. Notice that the above estimate follows a general calculation, with no restriction on the magnitude of $\epsilon$. This is at variance with the perturbative analysis that underlies the application of the averaging theorem and which eventually ensures the stability of the examined system for sufficiently small $\epsilon$ amount.

\section{Conclusion}

Synchronization plays a key role in many processes, either artificial or natural. In general terms, individual oscillators can be coupled via a complex web of interlinked connections, as specified by the architecture of an embedding network. Oscillators, sitting on the nodes of the network, are paired to their adjacent homologous, provided an edge exists which acts as a route for the spreading of the signal. When the coupling among distant sites senses the difference between respective dynamical variables, each anchored on one of the two interacting nodes (and assuming the oscillators to compose an ensemble made of identical units), a homogeneous synchronized solution exists which makes the network to oscillate in unison. Perfect synchronization can be however destroyed, upon injection of a small non homogeneous perturbation, for a specific set of reaction parameters and depending on the topological features of the underlying network. Working in this setting, we have here introduced and tested a control procedure, which aims at stabilizing the synchronous regime via network plasticity.  More concretely, we considered a family of Stuart-Landau oscillators, mutually coupled so as to turn the system unstable to an externally imposed perturbation. We then generated a second network,  tailored to the first, which links the same set of nodes. The second network is constructed so as to allow for synchronicity to be eventually restored, by iterating forward in time successive swaps between the two aforementioned networks, at a sufficient rate. As it can be intuitively grasped, the system framed on the time dependent support feels the average network, for frequent enough transitions between the two static networks. Stability is hence enforced by tuning the newly generated network, in such a way to have the system resilient to perturbation on the ensuing average support. The result is here made rigorous as follows an application of the average theorem, an approach that builds on the analysis of \cite{lucas}, where the dual scenario (inducing destabilization of the synchronous phase) was addressed. As a non trivial outcome of the theory, the critical value of the swapping frequency for securing synchronization is accurately estimated.


\begin{thebibliography}{99}


\bibitem{nicolis} G. Nicolis and I. Prigogine, {\it Self-organization in Nonequilibrium Systems}(Wiley, New York) 1977.

\bibitem{goldbeter} A. Goldbeter, {\it Biochemical OScillations and Cellular Rhythms} (Cambridge University Press,Cambridge) 1997.

\bibitem{pikovsky} A. M. Rosenblum and J. Kurths, {\it Synchronization: A Universal Concept in Nonlinear Sciences},Vol. {\bf 12} (Cambridge University Press,Cambridge) 2003.

\bibitem{strogatz} S. H. Strogatz, {\it Sync: The Emerging Science of Spontaneous Order}(Penguin, UK) 2004.

\bibitem{kuramoto} Y. Kuramoto, {\it Chemical Oscillations, Waves, and Turbolence}(Springer -Verlag, Tokyo) 1984.

\bibitem{barrat} A. Barrat, M. Barth\'elemy and A. Vespignani, {\it Dynamical processes on complex networks} (Cambridge University Press, Cambridge) 2008.

\bibitem{osipov} G. Osipov, J. Kurths and C. Zhou, {\it Synchronization in oscillatory networks} (Springer, Berlin, Germany) 2007.

\bibitem{barahona} M. Barahona and M. L. Pecora, {\it Phys. Rev. Lett.},{\bf 89} 054101 (2002).

\bibitem{arenas} A. Arenas, A. Daz-Guilera, J. Kurths, Y.Moreno and C. Zhou, {\it Phys. Rep.}, {bf 469} 93 (2008) .

\bibitem{kori} H. Kori and A.S. Mikhailov, {\it Phys. Rev. E},{\bf 74}  066115 (2006)

\bibitem{mirollo} S. H. Strogatz and R. E. Mirollo, {\it J. Stat. Phys.},{\bf 63}  613 (1991).

\bibitem{koseska} A. Koseska, E. Volkov and J. Kurths, {Phys. Rep.},{\bf 531}  173 (2013).

\bibitem{glass} L. Glass, {\it Nature}, {\bf 410}  27-284 (2001).

\bibitem{buck} J. Buck, {\it Q Rev. Biol.}, {\bf 63} 265-289 (1988).

\bibitem{dorfler} F. D\"orfler and F. Bullo, {\it SIAM Journal on Control and Optimization }, {\bf 50} 3 (2012) .

\bibitem{dorfler1} F. D\"orfler, M. Chertkov and F. Bullo, {\it PNAS}, {\bf 110}  6 (2013).


\bibitem{jung} D. Jung and S. Kettemann, {\it Phys. Rev. E},{ \bf 94}  012307 (2016). 

\bibitem{lucas} M. Lucas, D. Fanelli, T. Carletti, {\it Europhys. Lett.} {\bf 121}(5) 50008 (2018). 

\bibitem{julien} J. Petit, B. Lauwens, D. Fanelli and T. Carletti, {\it Phys. Rev. Lett.}, {\bf 119} 148301 (2017) .
 
 \bibitem{zoran} M. Faggian, F. Ginelli, F. Rosas, Z. Levnajic  {\it Scientific Reports}, {\bf 9}  10207 (2019). 

\bibitem{newman} M.E.J. Newman and D.J. Watts, {\it Phys. Lett. A },{\bf 263} (1999) 46 (1999).

\bibitem{cencetti} G. Cencetti, F. Bagnoli, G. Battistelli, L.Chisci, F. Di Patti and D. Fanelli, {\it Europ. Phys. J B}, {\bf 90} 9 (2017).

\bibitem{nakao} H. Nakao {\it Eur Phys J. Special Topics} {\bf 223}  2411-2421 (2014).

\bibitem{dipatti} F. Di Patti, D. Fanelli, F. Miele, T. Carletti {\it Chaos, Solitons and Fractals} {\bf 96} 8-16 (2017).


\end{thebibliography}
\end{document}